\documentstyle{article}
\title{\vspace*{-2.1cm}\hspace*{8.7cm} {\large gr-qc/9402034,
CGPG-94/2-2}\vspace*{1.45cm}\\
{\bf Is the exponential of the Chern-Simons action a normalizable
physical state?}}
\author{\\Guillermo A. Mena Marug\'an\vspace*{.6cm}\\
Center for Gravitational Physics and Geometry,\\ Pennsylvania State
University, 104 Davey Laboratory,\\ University Park, PA 16802, USA.
\vspace*{.4cm}\\ On leave from: {\it Instituto de Matem\'aticas y F\'{\i}sica
Fundamental},\\ {\it C.S.I.C., Serrano 121, 28006 Madrid, Spain.}\\
\vspace*{.4cm} }

\date{February, 1994}

\begin{document}
\renewcommand{\thefootnote}{\fnsymbol{footnote}}

\maketitle
\large
\setlength{\baselineskip}{.826cm}
\renewcommand{\headheight}{0cm}
\vspace*{.4cm}

\begin{center}
{\bf Abstract}
\end{center}

We determine the wavefunction that corresponds to
the exponential of the Chern-Simons
action in a family of gravitational minisuperspace models
provided with cosmological constant whose non-perturbative
canonical quantization is completely known.
We show that this wavefunction does not represent a proper
quantum state, because it is not normalizable
with respect to the unique inner product of Lorentzian gravity.

\vspace*{.8cm}

\newpage

The recent formulation by Ashtekar [1,2] of an alternative formalism
for the description of General Relativity has renewed the hopes of
constructing a consistent quantum theory of gravity by implementing
Dirac's canonical quantization program [3]. In addition to a change
of emphasis from geometrodynamics to connection dynamics, Ashtekar has
introduced a proposal to determine the inner product in the space of
solutions to all quantum gravitational constraints (the space of
physical states). One must find first a sufficient large number of
gravitational observables, that is, functions on phase space
that commute with all
the constraints. The inner product is then uniquely fixed [4] by
demanding that those observables that correspond to real classical
variables are represented in the quantum theory by self-adjoint
operators. These self-adjointness requirements are usually
called reality conditions [2].

Among the different achievements reached so far in the Ashtekar
formalism,
one of the more significant successes has been the discovery, for
the first time, of an exact solution to the quantum constraints of
full General Relativity. This physical state is provided by the
exponential of the Chern-Simons action [5], and its explicit form has
been obtained both in the connection [6,7] and in the loop
representation [5] of quantum gravity constructed using
Ashtekar variables [2,8]. Since this solution
(the Chern-Simons solution, from now on)
is the only exact state that is known for full gravity
with a non-vanishing cosmological term, it would be
extremely interesting to determine whether it belongs to
the Hilbert space of physical states for Lorentzian General Relativity,
i.e., if it actually represents a quantum gravitational state of
finite norm. In the present circumstances,
however, it seems impossible to arrive at a definitive
conclusion about the normalizability of the Chern-Simons
solution, because the expression of the inner product for
Lorentzian gravity has not been found yet.
A partial analysis can none the less be carried out by
restricting our discussion to gravitational minisuperspace models
provided with a cosmological constant for which both the physical
inner product and the Chern-Simons wavefunction can be obtained.

A lot of attention [6,9] has been devoted lately to the quantization
of minisuperspace models as a powerful tool to check the consistency of
the quantization program proposed by Ashtekar, as well as to
develop the mathematical machinery that will be presumably needed
to quantize the full theory of gravity. There exist several examples
in the literature in which the non-perturbative canonical quantization
has been performed to completion, including the determination of
the inner product [10,11]. In all these examples,
nevertheless, the cosmological constant
has been assumed to vanish,  except (to our knowledge) in the canonical
quantization of a family of anisotropic models
that contains the locally rotationally symmetric (LRS) Bianchi types I
and III and the Kantowski-Sachs model as particular cases [11,12].
Our aim in this letter is to find the explicit form of the
Chern-Simons solution in this family of minisuperspaces and show that,
with the inner product that corresponds to Lorentzian gravity,
such a wavefunction possesses an infinite norm, so that it
cannot be considered a proper quantum state.

The gravitational models that we are going to analyse can be
described by the spacetime metric [12]:
\begin{equation} ds^2=-\frac{\bar{N}^2(t)}{a^2(t)} dt^2+a^2(t) dr^2+
b^2(t) d\Omega^2_2.\end{equation}
Here, the coordinate $r$ is periodic with period equal
to $2\pi$, $a$ and $b$ are the two scale factors of the model, $\bar{N}$
is the rescaled lapse function, and $d\Omega^2_2$ denotes the metric of a
compact orientable two-manifold with constant scalar of curvature equal to
$2k$, $k=+1$, $0$ or $-1$, and volume given by $V_2$. The two-metric
$d\Omega^2_2$ can always be written locally as
\begin{equation} d\Omega^2_2=d\alpha^2+f^2(\alpha)d\beta^2,\end{equation}
where $f(\alpha)$ satisfies the differential equation
\begin{equation} \partial^2_{\alpha}f(\alpha)=-k f(\alpha).\end{equation}
The specific minisuperspace considered depends thus on the
value taken by $k$. For $k=+1$ we get the Kantowski-Sachs
model, while for $k=0$ and $-1$ we obtain, respectively, the LRS Bianchi
types I and III [12].

{}From now on, we will adopt the system of units in which
$\hbar=1$, $8\pi G=1$, and $4\pi V_2=1$,
$G$ being the gravitational constant.
The Hamiltonian constraint of the minisuperspaces (1), including the
contribution of the cosmological constant $\lambda$, can then
be expressed in the form [11,12]:
\begin{equation} {\cal H}=\frac{1}{2}(-4p_cp_b+\lambda b^2
-k)=0,\end{equation}
with $p_b$ and $p_c$ the momenta canonically conjugate to $b$ and
\begin{equation} c=a^2b.\end{equation}
The quantization of these models was first analysed by Halliwell and Louko
by using the complex path-integral approach in the geometrodynamic
formulation [12]. The complete non-perturbative canonical quantization
of these systems was presented in Ref. [11], where the unique inner
product compatible with the reality conditions of Lorentzian gravity was
determined.

Since we are interested in studying the behaviour of the Chern-Simons
wavefunction, which is originally defined in the connection
representation of the Ashtekar formulation [5-7], it will be useful to
begin our discussion by introducing the Ashtekar variables for the models
under consideration. Once we have found the explicit form of the
Chern-Simons solution, we will translate our results into the representation
that was used in [11] to achieve the canonical quantization, i.e., the
representation in which we know the physical inner product.

The Ashtekar variables are a densitized triad, $\tilde{E}^a_i$, and a
conjugate SO(3) connection, $A^i_a$ [1,2]. The lower case Latin
letters from the beginning and the middle of the alphabet denote here
spatial and SO(3) indices respectively, the latter being raised and
lowered with the metric $\eta^{ij}=(1,1,1)$. For non-degenerate metrics,
the Ashtekar variables can be obtained from the inverse triad,
$e^a_i$, and the extrinsic curvature, $K_{ab}$,
\[ \tilde{E}^a_i=e^a_i\,q(e),\;\;\;A^i_a=\Gamma^i_a(e)-iK_{ab}e^b_i,\]
$q(e)$ being the determinant of the triad and $\Gamma^i_a(e)$ the
SO(3) connection compatible with the three-metric [2]:
\[ \Gamma^i_a(e)=-\frac{1}{2}\epsilon^{ijk}E_{_{_{\!\!\!\!\!\sim}}\;jb}
(\partial_a\tilde{E}^b_k+\Gamma^b_{\;ca}\tilde{E}^c_k).\]
In this formula, $\epsilon^{ijk}$ denotes the antisymmetric symbol,
$\Gamma^a_{\;bc}$ the Christoffel symbol [13], and
$E_{_{_{\!\!\!\!\!\sim}}\;a}^i$ the inverse of $\tilde{E}^a_i$.

Using the coordinatization of (1,2) and imposing a convenient gauge
condition on the SO(3) degrees of freedom (so that the densitized triad be
diagonal), the particularization of
these equations to our minisuperspace models leads to the expressions:
\begin{equation} \tilde{E}^1_1=f(\alpha)\,x,\;\;\;\tilde{E}^2_2=f(\alpha)
\,y,\;\;\;\tilde{E}^3_3=y,\end{equation}
\begin{equation} A^1_1=-2iP_x,\;\;\;A^2_2=-iP_y,\;\;\;A^3_3=-if(\alpha)\,
P_y,\end{equation}
\begin{equation}A^1_3=\partial_{\alpha} f(\alpha),\end{equation}
where we have introduced the new variables
\begin{equation} x=b^2,\;\;\;y=ab,\end{equation}
and $P_x$ and $P_y$ are the momenta canonically conjugate to $x$ and $y$.
To obtain eq. (7) we have used the equations of motion derived
from the Hamiltonian (4). The rest of components of the densitized triad
and the Ashtekar connection vanish in the gauge selected here.

{}From eqs. (6-8), it is clear that all the degrees of freedom of the
Ashtekar variables reduce in this family of homogeneous models to the set of
functions $x$, $y$, $P_x$ and $P_y$. On the other hand, only the Hamiltonian
constraint remains to be imposed in these minisuperspaces, since we have
already fixed the diffeomorphism and SO(3) gauge freedom. Substituting then
eqs. (6-8) and eq. (3) in the general formula
for Ashtekar's Hamiltonian  constraint [2,14]:
\begin{equation} {\cal H}^{^{\!\!\!\!\!\sim^{\!\!\!\!\!\,\sim}}}=
\frac{1}{2}\epsilon^{ijk}\eta_{_{_{_{\!\!\!\!\!\sim}}}abc}
\tilde{E}^b_j\tilde{E}^c_k\,(\tilde{\cal F}^a_i
+\frac{\lambda}{3}\tilde{E}^a_i)=0,\end{equation}
\begin{equation} \tilde{\cal F}^a_i=\tilde{\eta}^{abc} (\partial_b A_{ci}
+\frac{1}{2}\epsilon_{ijk}A^j_b A^k_c)\end{equation}
(with $\tilde{\eta}^{abc}$ the Levi-Civitta tensor-density) and integrating
the result, divided  by $f(\alpha)$ \footnote{ Notice that
$f(\alpha)\neq 0$ for non-degenerate metrics (1,2).}, over each
constant time surface, we
arrive at a Hamiltonian of the form
\begin{equation} \bar{\cal H}=\frac{1}{2}(-4 xy P_x P_y- y^2 P_y^{\,2}
+\lambda xy^2-ky^2)=0.\end{equation}
For consistency in our calculations [2,14], this constraint must
coincide with eq. (4) when multiplied by the time dependent part of the
square root of the determinant of the three-metric
(note that we have integrated over the spatial dependence to derive eq. (12))
and by a factor of $a$ that arises from the definition of the rescaled
lapse function $\bar{N}$ in (1). That this is
indeed the case can be checked by using eqs. (5) and (9) and the relations
between the momenta $(P_x,P_y)$ and $(p_b,p_c)$ that follow from them:
\[ \frac{1}{y}P_y=\frac{2}{b}p_c,\;\;\;P_x=\frac{1}{2b}p_b-\frac{c}{2b^2}p_c.
\]

The exponential of the Chern-Simons action is known to be a
physical state in the connection representation of the Ashtekar
formulation only if all the Ashtekar connections are gathered to the
right of the densitized triads in the gravitational constraints before
quantization [5,6]. Under reduction to our minisuperspace
models, this prescription gives exactly the factor ordering that
appears in eq. (12), and the analogue of the connection representation
is provided now by the $(P_x,P_y)$ representation,
in which the operators $\hat{x}$ and $\hat{y}$ act as derivatives
with respect to their conjugate momenta
($\hat{x}=i\partial_{P_x}$ and $\hat{y}=i\partial_{P_y}$), and
$\hat{P}_x$ and $\hat{P}_y$ are multiplicative operators.

The Chern-Simons wavefunction is in fact a quantum solution to a more
restrictive equation than constraint (10), i.e.,
\begin{equation} \tilde{E}^a_i=-\frac{3}{\lambda}\tilde{\cal
F}^a_i.\end{equation}
This condition, together with the factor ordering commented
above, guarantees that the exponential of the Chern-Simons action is
annihilated by the quantum Hamiltonian obtained from eq. (10). Moreover,
when the topology of the constant time sections is fixed (like,
for instance, in the models that we are studying), the
Chern-Simons solution is characterized, up to a constant,
as the only wavefunction that satisfies the quantum version
of relation (13) [7]. For the family of
minisuperspaces analysed here, eqs. (3), (6-8) and (11)
allow us to rewrite that relation as the following pair of identities:
\begin{equation} x=\frac{3}{\lambda} (k+P_y^2),\;\;\;y=\frac{6}{\lambda}
P_xP_y.\end{equation}
A straightforward calculation shows then that, in the $(P_x,P_y)$
representation, the only quantum solution to eqs. (14) is provided by
\[ \Psi_{{\rm CS}}(P_x,P_y)=\exp\left(-i\frac{3}{\lambda}[kP_x+P_xP_y^2]
\right),\]
which must be therefore the Chern-Simons wavefunction.

In order to transcribe these results into the representation
in which the physical inner product is known [11],
let us assume first that there exists a well-defined $(x,y)$
representation for the quantum
theory with Hamiltonian constraint (12).
In such a representation, $\hat{P}_x=-i\partial_x$,
$\hat{P}_y=-i\partial_y$ and
$\hat{x}$ and $\hat{y}$ act as multiplicative operators.
Recalling eqs. (5) and (9),
we can define then an equivalent $(b,c)$ representation through the
following rescaling of the wavefunctions $\Psi(x,y)$,
\begin{equation} \Psi (b,c)=\frac{1}{\sqrt{b}}\Psi(x=b^2,y=\sqrt{cb}).
\end{equation}
Applying now the chain rule in the quantum Hamiltonian constraint
obtained from eq. (12), we conclude that
\begin{equation} \hat{\bar{{\cal H}}\;}\Psi(x,y)=\frac{cb\sqrt{b}}{2}(
-4\hat{p}_b\hat{p}_c+\lambda b^2-k)\Psi(b,c)=0,\end{equation}
where $\hat{p}_b=-i\partial_b$ and $\hat{p}_c=-i\partial_c$. Therefore,
if we neglect the prefactor $cb\sqrt{b}$ in the above equation, our
previous assumption about the viability of a $(x,y)$ representation
amounts to admit the existence of a $(b,c)$ representation for those
models whose only quantum constraint is precisely the Wheeler-DeWitt
equation that follows from eq. (4). That such a representation
exists was proved in fact in Ref. [11], where we showed that,
in the minisuperspaces with Hamiltonian (4), the
real $(b,c)$ representation is well-defined at least
for all physical states in the Hilbert space of Lorentzian gravity.
Thus, we can always adopt this $(b,c)$ representation to describe
the quantum systems under consideration.

The global factor $cb\sqrt{b}$ that appears in (16) actually
allows the wavefunctions $\Psi(b,c)$ to be Green functions of the
Wheeler-DeWitt equation associated with eq. (4), rather than mere
solutions. Eq. (16) is then satisfied in the distributional sense.
We will return to this point later in this letter.

Let now $\Psi_{CS}(b,c)$ be the Chern-Simons wavefunction in the
$(b,c)$ representation introduced above. We know that, in the models
that we are discussing, this wavefunction must exist, at least, if
the Chern-Simons solution is a normalizable physical state for
Lorentzian gravity. A straightforward calculation
employing eqs. (5), (9) and the chain rule shows then
that the quantum version of eqs. (14),
which determine the Chern-Simons wavefunction,
translate in the $(b,c)$ representation into
\begin{equation} \sqrt{b}\left(k-\frac{\lambda}{3}b^2+\frac{4c}{b}\hat{p}_c
^{\,2}-\frac{2i}{b}\hat{p}_c\right)\Psi_{CS}(b,c)=0,\end{equation}
\begin{equation}\frac{\sqrt{cb}}{b^2}\sqrt{b}\left(\hat{p}_b\hat{p}_c
-\frac{\lambda}{6}b^2-\frac{c}{b}\hat{p}_c^{\,2}+\frac{i}{2b}\hat{p}_c
\right)\Psi_{CS}(b,c)=0.\end{equation}

We are at last in an adequate position to rewrite our expressions
in the representation that was used in Ref. [11] to achieve the
non-perturbative canonical quantization of these systems. This quantization
was obtained by means of a transformation to a new set of canonical
variables, $(Q,P,H,T)$, related to the original set $(b,p_b,c,p_c)$
through the equations\footnote{ The first equation in (20) corrects a
missprint in eq. (51.c) of Ref. [11].}
\begin{equation} b=4PT,\;\;\;\;\;\;\;\;\;\;\;\;\;\;
p_b=4\lambda PT^2-\frac{1}{4P} H,\end{equation}
\begin{equation} c=Q+\frac{16}{3}\lambda PT^3-\frac{1}{P}(HT),\;\;\;\;
p_c=P.\end{equation}
$P$ and $Q$ are a canonically conjugate pair of observables (they
commute with the Hamiltonian constraint of these models), $T$ plays the
role of an intrinsic time, and $H$ is essentially the Hamiltonian constraint,
which now reads $H=k$ [11].
In the associated $(P,H)$ representation, in which $\hat{Q}=i\partial_P$
and $\hat{T}=-i\partial_H$, the physical states of the quantum theory are
provided by wavefunctions (distributions) of the form
\begin{equation} \Psi(P,H)=f(P) \delta(H-k).\end{equation}
The reality conditions that correspond to Lorentzian gravity fix
then the unique inner product [11]
\begin{equation} <\Phi,\Psi>=\int_{I\!\!\!\,R}dP\,h^{^{^{\!\!\!\!\!-\!\!-}}}
(P)f(P),\end{equation}
where $\Phi(P,H)=h(P)\delta(H-k)$ and $^{^{^{\!-\!\!-}}}$ denotes
complex conjugation. As a consequence, the Hilbert space of
quantum states turns
out to be simply the space of functions $f(P)\in L^2(I\!\!\!\,R)$.

In the adopted $(P,H)$ representation, and according to Ref. [11],
the action of the operators $(\hat{b},\hat{p}_b,\hat{c},\hat{p}_c)$
is given by the direct translation of eqs. (19,20) into operator language,
with the only caveat that the factor $(HT)$ that appears in the definition
of $c$ must be taken as a symmetric product. With this prescription
(and neglecting again algebraic factors that never vanish when $c$ and $b$
are different from zero), the quantum equations (17) and (18) that
characterize the Chern-Simons solution can be rewritten, respectively, as
\begin{equation} \left[(k-\hat{H})+(\hat{P}\hat{T})^{-1}(-i\hat{P}+\hat{Q}
\hat{P}^2)\right]\Psi_{CS}(P,H)=0,\end{equation}
\begin{equation} (-i\hat{P}+\hat{Q}\hat{P}^2)\,\Psi_{CS}(P,H)=0
.\end{equation}
Notice that eqs. (23,24) imply, in particular, that the Chern-Simons
solution satisfies the Hamiltonian constraint, $(\hat{H}-k)\Psi_{CS}=0$.
On the other hand, the presence of the singular operator
$(\hat{P}\hat{T})^{-1}$ in eq. (23) poses no difficulties provided that eq.
(24) is satisfied exactly. A trivial computation leads then
to the explicit form of the Chern-Simons wavefunction (up to a constant) in
the $(P,H)$ representation
\begin{equation} \Psi_{CS}(P,H)=f_{CS}(P)\delta(H-k),\;\;\,f_{CS}(P)=
\frac{1}{P}.\end{equation}
This wavefunction is obviously not normalizable with respect to
the inner product (22),
for $f_{CS}(P)=1/P$ is not square integrable over the real axis. We thus
conclude that the exponential of the Chern-Simons
action is a physical state which, at least in the minisuperspaces
that we are analysing, does not belong to the Hilbert space of quantum
Lorentzian gravity.

We had commented above that, in the $(b,c)$ representation,
the Ashtekar Hamiltonian constraint (16) admits also as solutions
(in the distributional sense) the Green functions of the Wheeler-DeWitt
equation that follows from eq. (4). In Ref. [11] we proved that, for
Lorentzian gravity, these Green functions can be obtained
from the discussed $(P,H)$ representation by restricting the domain of
the variable $P$ to be the real positive axis. With this restriction,
the normalizable physical states are still of the form (21), but now
$f(P)$ must belong to $L^2(I\!\!\!\,R^+)$. Hence,
the Chern-Simons wavefunction (25) turns out to possess an infinite norm
also in this case.

In conclusion, we have proved that the exponential of the Chern-Simons
action does not provide an acceptable quantum state in the Lorentzian
theories obtained from the non-perturbative canonical quantization of
the anisotropic models (1). There are only two reasons
that can prevent us from extending this result to the Ashtekar formulation
of full Lorentzian gravity. The first caveat refers to the particular
quantization that has been used in the analysis of the minisuperspace
models, quantization that was obtained by assuming that the variables
$b$ and $c$ run over the real axis [11]. For generic real domains
of the Ashtekar variables $x$ and $y$, however, the phase space coordinates
$b$ and $c$ may be complex in general, because the relations (5,9)
between these two sets of functions are not linear. It might then
happen that, for reality conditions that allow complex domains
for the classical variables $b$ and $c$ (so that $(x,y)$, and hence
the densitized triad (6), are real), there exists a different canonical
quantization in which the Chern-Simons wavefunction turned out to be
normalizable. The second and most important reason
is the drastic reduction of degrees of freedom that leads from full
General Relativity to the studied family of minisuperspaces.
This reduction may have significant physical consequences. In particular,
the displayed non-normalizability of the Chern-Simons solution might
simply be an artifact of the minisuperspace approximation analysed in
this work.

In spite of these open problems, our results clearly indicate that
one should not expect the exponential of the Chern-Simons action
to be a proper quantum state in full Lorentzian gravity. It
therefore seems that we do not know yet any exact solution to the quantum
gravitational constraints which, in the presence of a cosmological
constant, can describe a truly physical state.

\vspace*{.2cm}
This work was supported by funds provided by
the Spanish Ministry of Education and Science Grant No. EX92-06996911.

\end{document}